\documentclass[journal=jacsat,manuscript=article]{achemso}

\usepackage{chemformula} 
\usepackage[T1]{fontenc} 
\usepackage{multirow}
\usepackage{amssymb}
\usepackage{amsmath}
\usepackage{tikz}
\usepackage{hyperref}
\usepackage{cleveref}
\usetikzlibrary{bayesnet}

\author{Benson Chen}
\affiliation[Insitro]
{Insitro, South San Francisco, California 94080, United States}
\email{bensonc@insitro.com}
\author{Mohammad M. Sultan}
\affiliation[Insitro]
{Insitro, South San Francisco, California 94080, United States}
\email{msultan@insitro.com}
\author{Theofanis Karaletsos}
\affiliation[Insitro]
{Insitro, South San Francisco, California 94080, United States}
\email{theofanis@insitro.com}

\title{Compositional Deep Probabilistic Models of DNA-Encoded Libraries}
\abbreviations{IR,NMR,UV}
\keywords{American Chemical Society, \LaTeX}

\begin{document}
\newpage


\begin{abstract}
  DNA-Encoded Library (DEL) has proven to be a powerful tool that utilizes combinatorially constructed small molecules to facilitate highly efficient screening experiments. These selection experiments, involving multiple stages of washing, elution, and identification of potent binders via unique DNA barcodes, often generate complex data. This complexity can potentially mask the underlying signals, necessitating the application of computational tools such as machine learning to uncover valuable insights. We introduce a compositional deep probabilistic model of DEL data, {\bf DEL-Compose}, which decomposes molecular representations into their mono-synthon, di-synthon, and tri-synthon building blocks and capitalizes on the inherent hierarchical structure of these molecules by modeling latent reactions between embedded synthons. Additionally, we investigate methods to improve the observation models for DEL count data such as integrating covariate factors to more effectively account for data noise. Across two popular public benchmark datasets (CA-IX and HRP), our model demonstrates strong performance compared to count baselines, enriches the correct pharmacophores, and offers valuable insights via its intrinsic interpretable structure, thereby providing a robust tool for the analysis of DEL data.
\end{abstract}

\newpage

\section{Introduction}

\begin{figure}
    \centering
    \includegraphics[width=\textwidth]{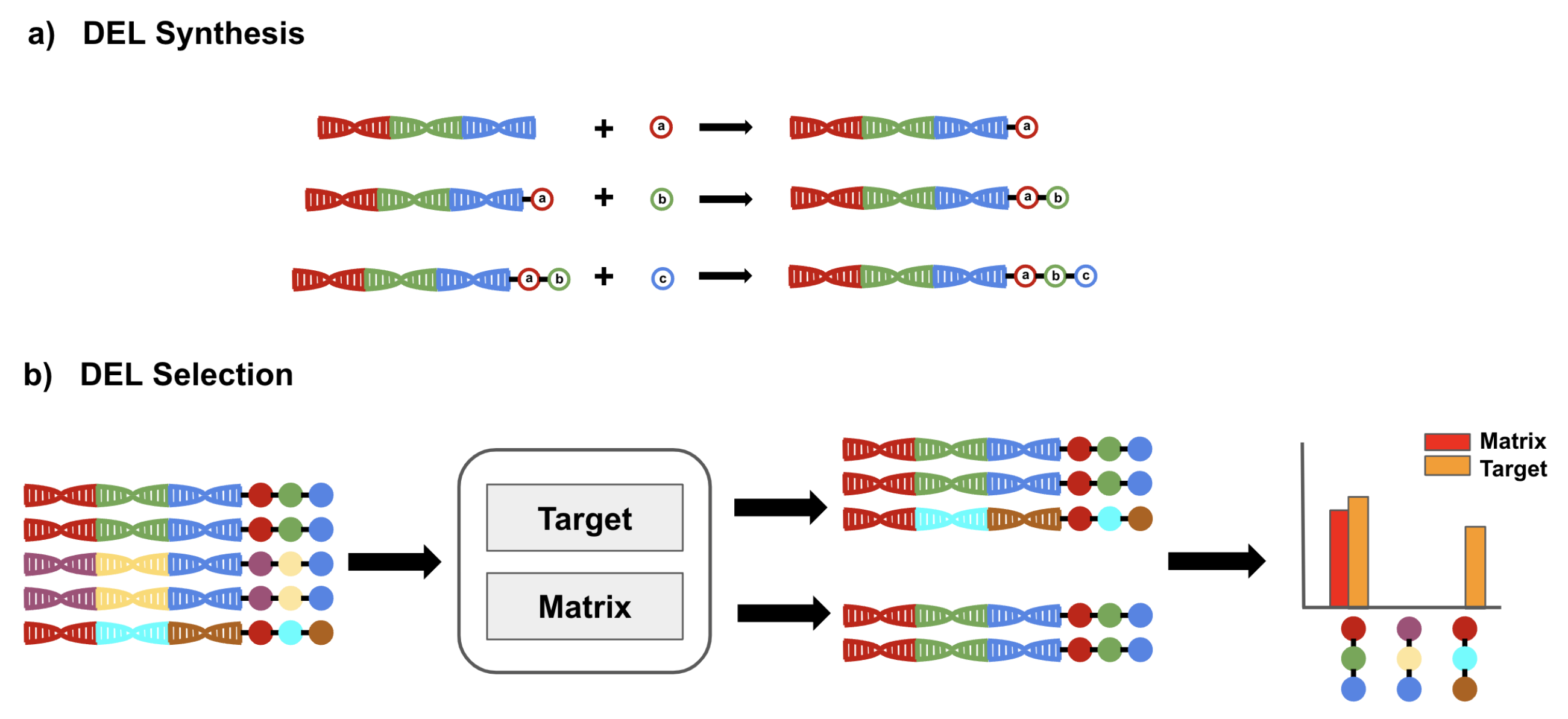}
    \caption{a) DELs are synthesized in a sequential manner via a split-and-pool method. The DNA barcode encode the identity of the synthon building block that should be added at each step. b) The pool of DEL molecules is then used in selection experiments that washes off any weak binders both in the presence of the protein of interest (\textbf{target}) and in the absence of the protein (\textbf{matrix}). The DNA of surviving library members are then amplified and sequenced to get count data correlated to likely binders.}
    \label{fig:intro_fig}
\end{figure}

DNA-Encoded Libraries (DELs) have demonstrated their potency as a robust method for conducting efficient exploration across a vast chemical landscape, and has recently gained significant traction in drug discovery efforts~\cite{goodnow2017dna,yuen2017achievements,neri2018dna,madsen2020overview, satz2022dna,peterson2023small}. These small molecule libraries are synthsized combinatorially by combining diverse building blocks with compatible chemistries. A DNA barcode, which is covalently attached to the molecule, specifies the unique combination of building blocks for each molecule. DELs are then used in selection experiments for proteins of interest, wherein multiple rounds of washing and elution are performed before identification of the surviving library molecules. We briefly illustrate the process in Figure~\ref{fig:intro_fig}. While proven to be a highly efficient process of exploring chemical space at scale, these selection experiments are noisy and require computation methods with the correct inductive biases to extract useful signals for downstream applications such as hit discovery and lead optimization~\cite{gironda2021dna, reiher2021trends}.

\paragraph{Prior work has tackled the analysis of DEL data from various perspectives.} Many of these methods are predicated on computing an enrichment score for each molecule as a function of the observed data or molecule structure. \citet{gerry2019dna} computes this enrichment score by fitting Poisson distributions to the count data and then computing the ratio of the on-target versus off-target binding events derived from the fitted distributions. However, this approach and other similar approaches~\cite{kuai2018randomness, faver2019quantitative} do not extend to out-of-library predictions as the enrichment is computed from the data itself rather than predicted via molecular structure. To that end, other methods have tackled this problem by utilizing molecule structure via molecular fingerprints and graph neural networks. \citet{mccloskey2020machine} and \citet{zhang2023building} bin the count data and construct a classification problem based on their discretizations of the data. In particular, \citet{zhang2023building} proposes exploiting the compositional structure of DELs for the extraction of enrichment signals, but unlike our work, no explicit generative models of this factorized representation or the improved likelihoods to deal with DEL data are built.
Other approaches formulate the problem as a latent-variable prediction task, maximizing the probability of observing the count data under some prescribed probability distribution such as the Poisson or Negative Binomial distribution~\cite{binder2022partial, lim2022machine, ma2021regression}. \citet{shmilovich2023dock} extends the representation capabilities of models on DEL data by incorporating 3-D docked poses to enhance the performance of models without requiring additional supervised validation data. However, these prior works do not leverage the inherent hierarchical structure of DEL data. 

\paragraph{We introduce an new approach to modeling DEL molecules,} which explicitly factorizes the molecular representation in a motivated manner through the construction of a generative model. We propose learning individual synthon representations, and construct the corresponding di-synthon and tri-synthon representation from their respective synthon composition parametrized by neural networks. We use the term latent reactions to denote the construction of these embeddings, which is similar to how molecules are synthesized via chemical reactions, but in embedding space. Signals in DEL selection experiments are obfuscated by the various sources of noise, but given the combinatorial construction of the library, these sources of variation are highly correlated within any particular synthon group. For instance, PCR bias which can arise from specific codons will result in correlated biases for a particular synthon.
Naturally, this decomposition allows us to better attribute interpretability in our model, and our method additionally avoids the necessity of enumerating full-molecule structures, which requires careful specification of reaction templates which can be tedious and result in errors.

In addition to our newly proposed paradigm for representing DEL molecules computationally in this factorized fashion, we further investigate the effects of different experimental biases in order to aptly model the count data. In particular, we focus on two prominent sources of noise inherent in DEL data, which stem from pre-selection and replicate-level biases. Modeling these is typically omitted in previous work on this topic. Since DEL molecules are synthesized using a split-and-pool method, the relative abundance of each library member is uncertain in the final mixture. While the library itself is sequenced to obtain a rough estimate of the molecule distribution, this count data is also prone to potential synthesis and sequencing biases. Across different replicates, we also expect to see different experimental or sequencing noise. We propose a structured parametrization in our count likelihoods to account for the effects of these factors in order to better model the observed count data and learn useful latent properties of DEL molecules.

\paragraph{We test our models empirically on DEL selection datasets} for two targets: Carbonic Anhyrase IX (CA-IX) and Horseradish peroxidase (HRP)~\cite{gerry2019dna}. Since these two well-studied targets have known pharmacophores, we demonstrate that our model can effectively pick out the important synthons--even on challenging splits of the data. Furthermore, we demonstrate that our method can obtain competitive performance even without requiring fully enumerated molecule structures. Lastly, we show that our model offers useful insights into the predictions given by the model by assigning importance weights to synthons combinations via attention mechanisms, which correctly ranks synthon importance, and predicting possible dropout in the data to illustrate inferred noise.

\section{Methods}

\subsection{Molecules As Synthon Compositions}
Our model broadly capitalizes on the combinatorial nature of DEL molecules (visualized in Figure \ref{fig:intro_fig}), and creates a composition of representations using the individual building blocks of each molecule.
To that end, we describe a generative model of the underlying data-generating process for DEL count data and first introduce some mathematical notation.

Let $\mathcal{X}$ be the set of DEL molecules in our dataset, and $\{\mathcal{S}_A, \mathcal{S}_B, \mathcal{S}_C\}$ be the sets of synthons at the first, second and third positions respectively. Each molecule is denoted by $x_{abc} \in \mathcal{X}$, where the subscript indicates the identity of the synthon at a particular position ($a \in \mathcal{S}_A, b \in \mathcal{S}_B, c \in \mathcal{S}_C $). To simplify (and overload) notation, we omit the subscript for a particular synthon position if it is absent. For instance, $x_b$ denotes the molecule corresponding to the synthon $b$ at the second position, and $x_{ab}$ denotes the molecule corresponding to the combination of synthon $a$ at the first position and $b$ at the second position. 

DEL molecules are used in selection experiments wherein molecules undergo multiple rounds of washes to determine the strongest binders. Molecules with strong binding affinity would not be eluted off, but this binding might not be specific to the protein of interest. In order to also account for non-specific binding of the molecules, there are two typically two experimental conditions that are run, the \textbf{target} condition, which describes the data for selection against the protein target of interest, and the \textbf{matrix} condition, which describes the data in the absence of the protein target. Once the selection experiments are conducted, the surviving DEL members are sequenced, resulting in DNA read count data which we will denote as $C_t = \{c^i_t | i \in [1, n_t]\}$ and $C_m = \{c^j_m | j \in [1, n_m] \}$ for target and matrix read counts respectively. Here, $(n_t, n_m)$ are the number of count replicates for target and matrix respectively. Moreover, DEL data is usually calibrated with an additional read-out of the library itself, which we denote as $c_p$ (this notation is lowercase, as there is usually only a single read-out of the library). This library read-out is a noisy estimate of the relative abundance of each molecule member. 

\subsection{DEL-Compose: A Generative Model of DEL Data}
\label{sec:genMod}

\begin{figure}
    \includegraphics[width=0.55\textwidth]{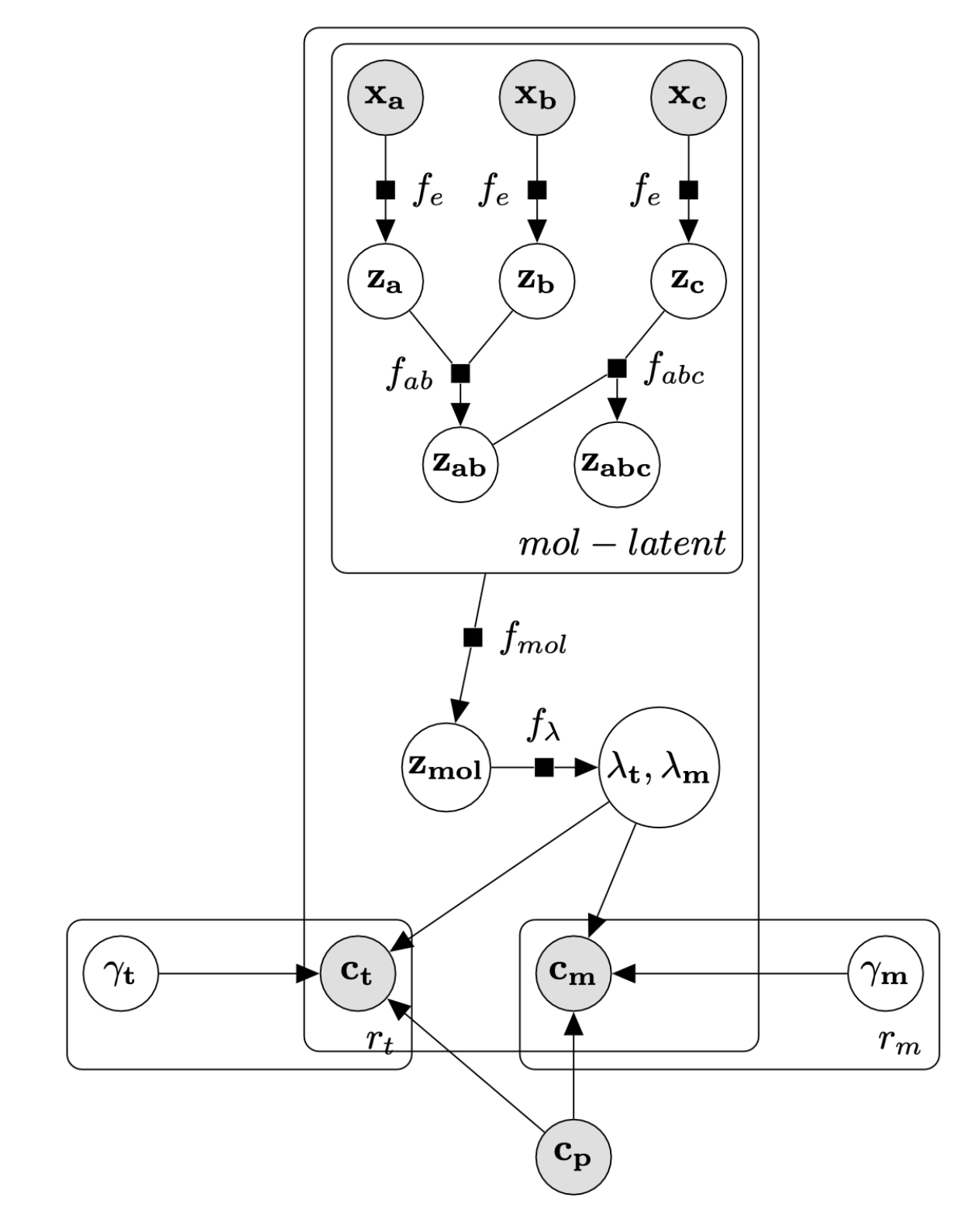}
    \caption{Graphical model depicting data-generating process for DEL count data. }
    \label{fig:plate}
\end{figure}

Here we introduce {\bf DEL-Compose}, a general paradigm for modeling DEL molecules leveraging their combinatorial nature (Figure \ref{fig:plate}), and we formulate the objective for a typical DEL design, but recognize that other variants exist. Our objective is to maximize the likelihood of observing the count data given an input molecule $x_{abc}$. Let $\mathcal{Z} = \{z_{s} 
 \in \mathcal{R}^{d}| s \in [a, b, c, ab, abc, mol] \}$ be a collection of latent synthon embeddings each of dimension $d$. $z_a$ denotes an embedding of an individual synthon $x_a$, while $z_{ab}$ denotes an embedding of a di-synthon $x_{ab}$ which is the reaction output of $x_a$ and $x_b$. Similarly, $z_{abc}$ denotes an embedding of tri-synthon $x_{abc}$, which itself is a product of the reaction of $x_{ab}$ and $x_{c}$. Finally, $z_{mol}$ is the aggregated embedding of all the above representations. We thus assume embeddings corresponding to all the (partial) products and building blocks of molecule can be viewed as a synthon decomposition and their respective set of chemical reactions. 
 
We utilize this structure to factorize a model of count observations given synthons $p(C_t, C_m | x_a, x_b, x_c)$ into two quantities as shown in Equation \ref{eq:prob_1}: a model capturing our beliefs about count observations given a collection of synthon embeddings $\mathcal{Z}$ and a model mapping observed synthons to such embeddings. Note that we do not require access to the full molecule observation $x_{abc}$, but rather only the individual synthons: $x_a$, $x_b$, and $x_c$ in such a model.

\begin{equation}
p(C_t, C_m | x_a, x_b, x_c) = \int p(C_t, C_m| \mathcal{Z}) \cdot p(\mathcal{Z} | x_a, x_b, x_c) d\mathcal{Z}.
\label{eq:prob_1}
\end{equation}

In practice we will be inferring point estimates of the embeddings corresponding to $p(\mathcal{Z}| x_a, x_b, x_c)$ but present the model in generality.
In order to de-noise the contribution of actual molecule binding to the count data read-outs, we explicitly define latent enrichment parameters $\{\lambda_t, \lambda_m\}$, which capture a molecule's affinity for binding in the target and matrix experimental conditions. While there are many auxiliary factors that affect the final read count for DEL experiments, we choose two prominent factors to incorporate in our model, which are pre-selection library read-out, $c_p$, and replicate-level noise, which we denote as $\{\gamma_t, \gamma_m\}$.
The generative model can then be broken down according to Equation \ref{eq:prob_2}:

\begin{equation}
p(C_t, C_m, \Phi_{mol},\mathcal{Z} | x_a, x_b, x_c; \Theta) = p(C_t, C_m | \Phi_{mol}; \Theta_x) p(\Phi_{mol} | \mathcal{Z}; \Theta_{o}) p(\mathcal{Z}| x_a, x_b, x_c; \Theta_i).
\label{eq:prob_2}
\end{equation}

Here, $\Phi_{mol}$ is the set of variables predicted by the model that parameterizes the output distribution, for instance $(\lambda_t, \lambda_m)$. The model is then broken up into three components: (1) $\Theta_i$ consists of the model parameters that construct the hierarchical synthon embeddings, (2) $\Theta_o$ consists of the model parameters that predict $\Phi_{mol}$ from these embeddings, and (3) $\Theta_x = \{c_p, \gamma_t, \gamma_m\}$ consists of the parameters for the observation model that captures experimental noise. We define the joint set of these parameters as $\Theta := \{\Theta_i, \Theta_{o}, \Theta_x \}$. We summarize the joint model visually in Figure~\ref{fig:plate} and will develop and define its components in more detail over the following sections.


\subsection{Neural Network Representations of synthons and molecules}

Here, we delienate our explicit modeling choices with respect to the probabilistic model introduced earlier, which is captured as the top part of the model visualized in Figure~\ref{fig:plate}.
Assuming $\mathcal{Z} = \{z_{s} | s \in [a, b, c, ab, abc, mol] \}$ we unpack $p(\mathcal{Z}| x_a, x_b, x_c; \Theta_i)$ to yield expressions in detail for each latent variable as follows:

\begin{equation}
p(\mathcal{Z}| x_a, x_b, x_c; \Theta_i)=\prod \limits_{s}^{\{a,b,c\}} \Bigl[ p(z_s|x_s;f_{e})\Bigr] p(z_{ab} | z_a, z_b; f_{ab}) p(z_{abc}|z_{ab}, z_{c}; f_{abc})p(z_{mol}|\mathcal{Z}_{\neq mol}; f_{mol}),
\end{equation}

where $\{f_{s}| s \in [a, b, c, ab, abc, mol]\}$ is a set of functions modeling the generative process by which the synthon embedding $z_s$ can be constructed, including (i) transformations from observed structures to embeddings and (ii) latent reactions modeling how synthon embeddings compose in latent space to form embeddings over higher order synthon structures.

\paragraph{Neural networks parametrize synthon embeddings}
We first identify function $f_{e}$ as a model parametrized by a multi-layer perceptron with parameters $\theta_e$ mapping molecular fingerprints $\phi(x_s)$ of a synthon $x_s$ to an embedding per synthon $z_s$.
 Let $\phi$ be a fingerprint transformation such as Morgan Fingerprints~\cite{rogers2010extended}; we compute the latent mono-synthon representations as $z_s = f_e \large(\phi(x_s);\theta_e \large)$. $\phi$ can also extend to other molecule representations such as graph neural networks (or other higher order functions that act on different data modalities of molecules), but we find fingerprints work well empirically and are fast to compute.

\begin{figure}
    \centering
    \includegraphics[width=0.95\textwidth]{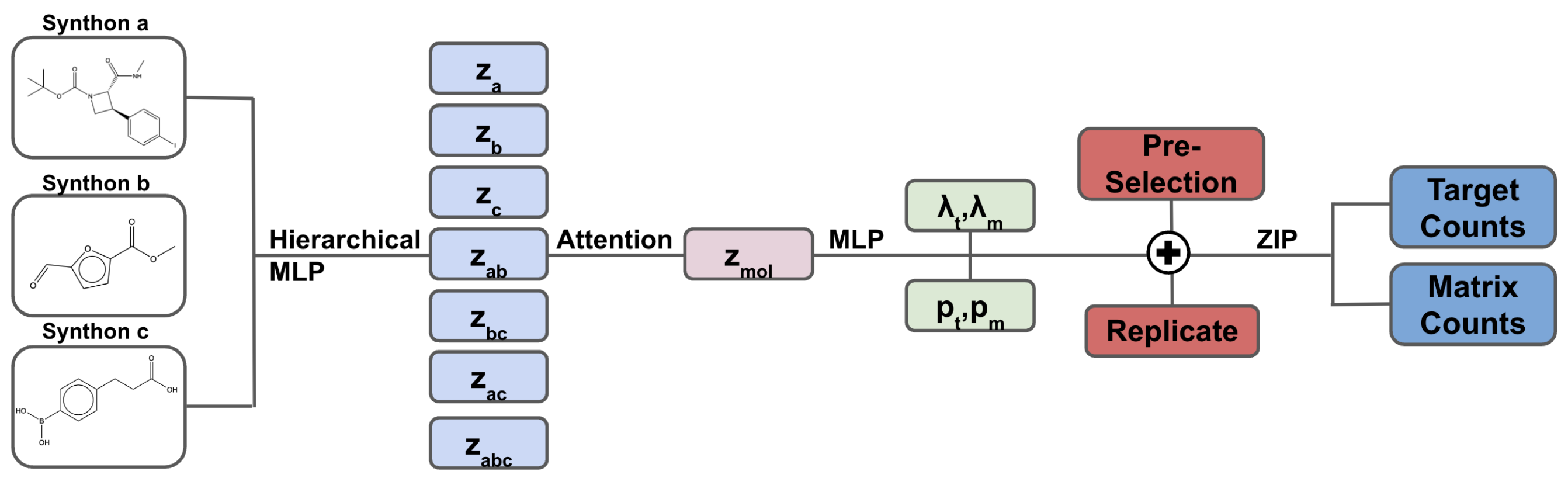}
    \caption{Model architecture for our model utilizing synthon-based embeddings. Synthons embedded and composed to mono-, di-, and tri-synthon embeddings, before aggregated into a single representation. Conditioned on the molecule embedding, our model predicts latent properties of the molecules that are used to model the observed count data. While the generative process does not explicitly incorporate $x_{bc}$ and $x_{ac}$, these are sensible molecular subgraphs that make sense in the representation of the molecule.}
    \label{fig:model}
\end{figure}

\paragraph{Neural networks parametrize latent reactions between synthon embeddings}
We then compute di-synthon and tri-synthon embeddings using mono-synthon embeddings, and give some examples as follows. Di-synthon embeddings utilize a latent reaction $f_{ab}$ to compute embedding $z_{ab} = f_{ab}([z_a, z_b];{\theta_{ab}})$ and tri-synthon embeddings are analogously given as $z_{abc} = f_{abc}([z_{ab}, z_c];\theta_{abc})$. Here, $(f_{ab}, f_{abc})$ are separate neural networks parameterized by $(\theta_{ab}, \theta_{abc})$ respectively. $f_{ab}$ can be interpreted to model latent reactions between mono-synthon embeddings $z_a$ and mono-synthon embeddings $z_b$ to yield a product-synthon-embedding $z_{ab}$. Likewise, $f_{abc}$ models the function mapping $z_{ab}$ and $z_{c}$ to the tri-synthon-embedding $z_{abc}$.
There are multiple ways to generate the tri-synthon embedding, for instance instead using all three mono-synthon embeddings, but we choose one formulation that is consistent with the sequential nature of a DEL molecule's synthesis. Although $(x_{bc}, x_{ac})$ are not actual observed partial products, we can optionally learn their respective embeddings and incorporate them into the model, as illustrated in Figure \ref{fig:model}. Finally, we utilize a multi-head attention layer $f_{mol}$ with parameters $\theta_{mol}$ on the different synthon embeddings to construct a single embedding representing the entire molecule, $z_{mol} = \textrm{Multihead-Attention}([z_a, z_b, z_c, z_{ab}, z_{bc}, z_{ac}, z_{abc}];\theta_{mol})$.
We note that $z_{mol}$ can be different from $z_{abc}$ or trivially equal to it. The model has the freedom to utilize $z_{mol}$ to focus on abstracting partial products if those are more informative about enrichment, while $z_{abc}$ is an explicit representation of the tri-synthon embedding and does not have to be maximally informative about enrichment on its own. We summarize all parameters of these individually specified functions as $\Theta_i = \{\theta_e, \theta_{ab}, \theta_{abc}, \theta_{mol} \}$.

\subsection{Probabilistic Models of Enrichment Counts}
\label{sec:likelihood}
In this section we will present some choices for the model generating the per-molecule observation model parameters given the embeddings $p(\Phi_{mol} | \mathcal{Z}; \Theta_{o})$.
We model the observed count as a zero-inflated Poisson (ZIP) distribution (PDF given in Equation \ref{eq:zip}), the parametrization of which is conditioned on our learned embeddings. A ZIP distribution is a mixture distribution that makes a bi-modal outcome assumption: a zero outcome indicating absence of measurable data, and a measurable enrichment with some appropriate count distribution. 
Intuitively, because DEL data is highly susceptible to noise of different types, including synthesis, amplification, or sequencing noise, we can think of the zero-probability as a drop-out parameter explaining away the absence of an expected count rather than forcing the model to absorb it by adjusting the expected enrichment.

\begin{equation}\label{eq:zip}
\text{P}(C=c | \lambda, \pi) = 
    \begin{cases}
        \pi + (1-\pi)e^{-\lambda} & \text{if } c = 0\\
        (1-\pi)\frac{\lambda^c e^{-\lambda}}{c!} & \text{if } c > 0\\
    \end{cases}
\end{equation}

For each of the experimental conditions, the target and matrix, we predict a separate ZIP with correlated parameters. A ZIP is characterized by two parameters, a mean value $\lambda$, and a zero probability $p \in [0, 1)$ that describes the rate of dropout. From our learned molecule embedding $z_{mol}$, we predict two sets of parameters $\{\lambda_t, p_t\}$ and $\{\lambda_m, p_m\}$ using the function $f_{\lambda}$ instantiated by an MLP with parameters $\theta_o$ (and consequently $\Theta_o := \theta_o$). For this likelihood this means that $\Phi_{mol} = \{\lambda_t, p_t, \lambda_m, p_m\}$, but we note that $\Phi_{mol}$ can take different appropriate shapes if other likelihoods are chosen.
: $[\lambda_t, p_t, \lambda_m, p_m] = f_{\lambda}(z_{mol};\theta_{o})$. $\lambda$ can be thought of as a molecule's intrinsic binding affinity property as learned by the model, while $p$ captures the noise in the data. 
This also highlights the utility of capturing enrichment-related molecule abstractions using $z_{mol}$, which can focus on utilizing partial products to predict noisiness and enrichment with more fidelity as our experiments will show.
If the data is inherently noisy and experience high dropout (or zero counts), the zero probability parameter should be relatively high. 

Since we expect that the binding in the matrix condition should be informative of a molecule's off-target binding, we want to correlate the predicted distribution of the target condition to the matrix. Additionally, we add the pre-selection counts $c_p$ and learned replicate-level effects $\{\gamma_t, \gamma_m\}$ as multiplicative factors in the mean of the predicted distribution. The latter accounts for variance across different replicates of the same experiment, which can result from PCR bias noise. We denote the collection of the relevant parameters for modeling experimental conditions and noise as $\Theta_x = \{c_p, \gamma_t, \gamma_m\}$. Together, we arrive at the following function form:

\begin{equation}
c^i_t \sim \textrm{ZIPoisson}\ (c_p \cdot \exp{(\gamma^i_t)} \cdot \exp{(\lambda_m + \lambda_t)}, p_t), \\
\end{equation}

\begin{equation}
c^j_m \sim \textrm{ZIPoisson}\ (c_p \cdot \exp{(\gamma^j_m)} \cdot \exp{(\lambda_m)}, p_m ),
\end{equation}
where $i$ and $j$ are target and matrix replicate indices, respectively. The output and count models introduced in this section are captured in the bottom part of the graphical model depicted in Figure~\ref{fig:plate}.

\section{Data}

We conduct experiments on public DEL data from \citet{gerry2019dna}, which includes DEL selection data on two well-studied protein targets: carbonic anhydrase IX (CA-IX) and horseradish peroxidase (HRP). This DEL is a tri-synthon library, consisting of 8 synthons at the A position, 114 synthons at the B position, and 118 synthons at the C position (107,616 total molecules) chosen to encourage chemical diversity of the library. Their data consists of two experimental conditions, one with the protein target, and one matrix condition that is conducted in the absence of the protein as control. For CA-IX this dataset includes 2 replicates of matrix data, and 4 replicates of on-target protein data; while for HRP, this dataset includes 4 replicates of matrix data and 2 replicates of on-target protein data. Additionally, there is data collected on the pre-selection library, which is an indicator of the relative abundance of the different DEL members.  

\begin{figure}
    \centering
    \includegraphics[width=0.78\textwidth]{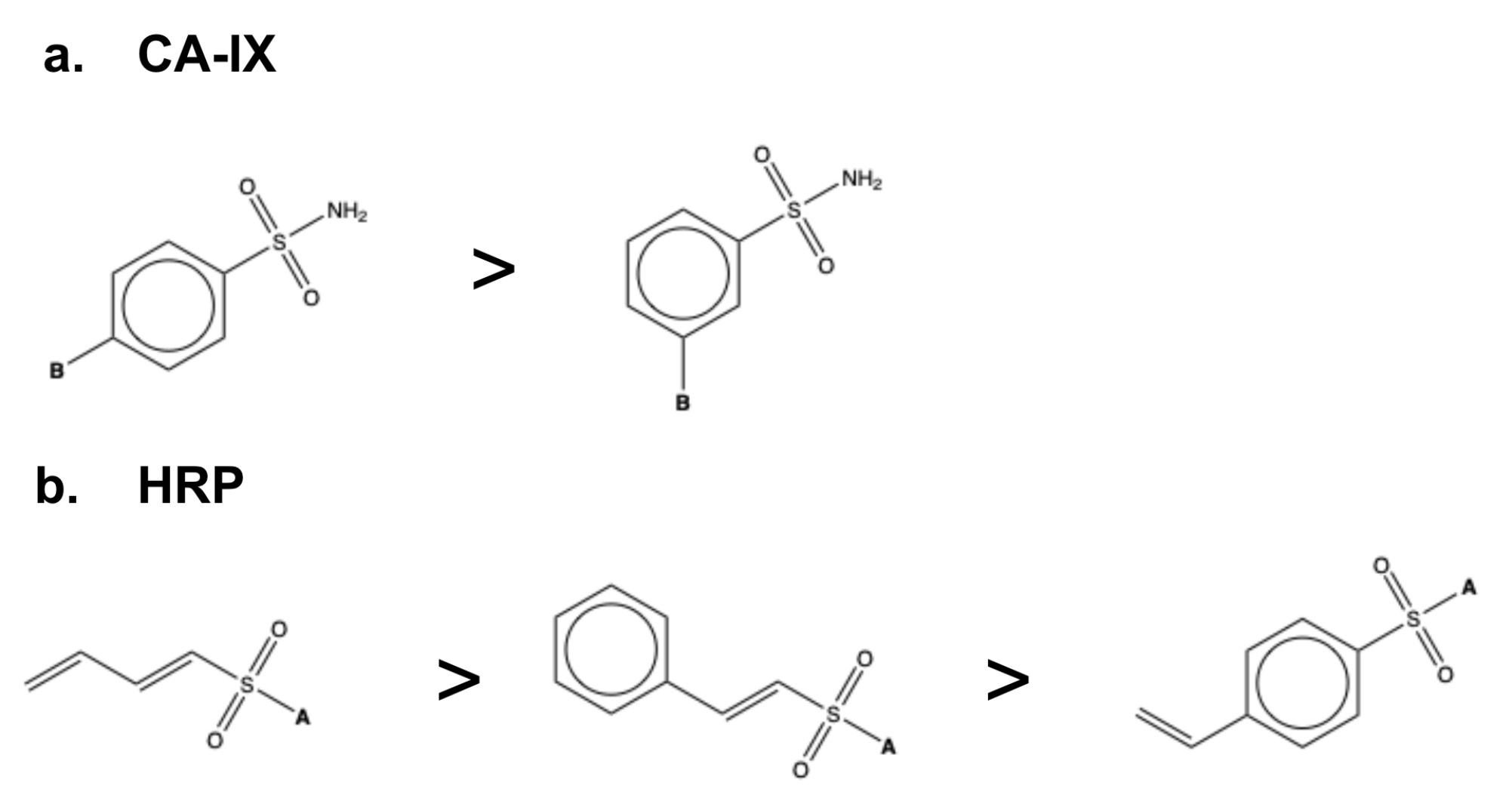}
    \caption{Known pharmacophores for both CA-IX (top) and HRP (bottom) pictured in descending order of affinity. For CA-IX, benzene-sulfonamides are known structures to induce affinity. The substitution of the sulfonamide affects the reactivity of the chemical specie, wherein the para-substituted constituent is found to be much more active. For HRP, electrophilic Michael acceptors are known pharmacophores. The structures are labeled with attachment to other synthons: for CA-IX, the sulfonamides are at the C positon with attachment to synthon B, and for HRP, the Michael acceptors are at the B position, with attachment to synthon A.}
    \label{fig:synthons}
\end{figure}

Both CA-IX and HRP are proteins with known pharmacophores~\cite{gerry2019dna}. CA-IX has a well-known binding motif, benzene-sulfonamide~\cite{buller2011selection}. In this dataset, there are two synthons at the C position that includes benzene-sulfonamides, one that is meta-substituted with respect to the aryl group, and the other which is para-substituted. Studies have shown that the benzene-sulfonamide substituted at the para position is much more highly active, in general, towards CA-IX protein~\cite{pagnozzi2022interaction}. Meanwhile, HRP is a enzyme with high affinity for compounds containing sulfonyl chloride–derived Michael acceptors~\cite{gerry2019dna, gan2013identification}. In this dataset, there are three such synthons at the B position that shows high activity, and are the three synthons we treat as ``gold'' labels for HRP. These structures are all visualized in Figure \ref{fig:synthons} in descending order of affinity.

\section{Results and Discussion}

For all results we present in the following, models are trained to minimize the negative log likelihood (NLL): $- \log[ p(C_t, C_m, \Phi_{mol},\mathcal{Z} | x_a, x_b, x_c; \Theta)]$ through gradient descent on parameters $\Theta$ using the Adam optimizer ~\cite{kingma2014adam}, which is a typical optimization technique that utilizes moving averages of gradients to help the model converge faster.

\subsection{Setup for in-distribution generalization}
In order to validate our model's performance, we propose a few different training scenarios. At the most primitive level, we want to evaluate our model's performance on a held-out test set of the data. To that end, we randomly split the data into 5 different splits of 80\%/10\%/10\% for train/validation/test sets respectively. Models are trained on the train set, selected based on the validation set and then finally tested on the held-out test set. Where applicable, our results are averaged across the 5 different splits, and we report the standard deviation across the splits.

\subsection{Setup for out-of-distribution generalization}
Random splits are not always ideal for testing molecule datasets~\cite{gawlikowski2023survey, heid2023characterizing}. In order to test the generalizability of molecule representations, many approaches attempt to split molecule by molecular scaffolds~\cite{wu2018moleculenet}. For DELs, rather than generic molecular scaffolding strategies, synthons provide a natural grouping and separation of the chemical space. By using synthons to split the data, we can test the generalizability of the model on unseen chemical structures. In the dataset that we are using, the known pharmacophores are conveniently localized to specific synthons, so we can develop intuitive splitting strategies. Most of the signal is captured by these pharmacophores, so we cannot withhold all of these molecules from training. Instead, we split on the synthon position that does not include these individual pharmacophores. Specifically, for CA-IX, the benzene-sulfonamides are at the C position, so we create synthon splits by splitting on the B position. For HRP, the electrophilic Michael acceptors are at the B position, so we split the data at the C position.

\subsection{Setup for data-efficiency generalization}
To understand more about our models, we introduce a third setup that tests the ability of the model to adapt under low-resource regimes. Since most of the signal resides in the molecules with known pharmacophores for their respective targets, we investigate the performance of our model when we change the amount of data provided to it. This will allow us to determine the quantity of data required to learn a reasonable model. In particular, these experiments provide a good way to compare different representational modalities, as we expect that our factorized approach should learn faster under resource-limited regimes.

\subsection{Metrics}
We utilize several well-motivated metrics to evaluate the performance of our model without additional data (ie on-DNA $K_D$ data for DEL molecules). Since we model the observed data through predicting a count distribution, we can measure the performance through the model loss, which is the negative log likelihood (NLL) of the ZIP distribution predicted for molecules in a held-out test set. This is a typical metric to gauge the overall fitness of a probabilistic model. However, as with other applications, likelihood metrics can be complemented with other application-relevant metrics to capture the behavior of the model.

Since we are interested in the quality of the learned model, we want to directly capture its ability to represent signals in the data. We use the expected mean of the predicted distribution as the computed enrichment, or affinity, of a molecule. For our model, this is exactly $\epsilon = (1- p) \cdot \lambda$, where $p$ is the predicted zero-probability and $\lambda$ is the predicted latent score for a molecule. While DEL-Compose predicts two count distributions, one for the target and one for the matrix, the latter is mainly used to calibrate a molecule's affinity for the protein target. A molecule with high counts in the matrix but not the target condition should be predicted to have a high matrix enrichment score, but a low target enrichment score. Using these enrichment scores, we can gauge the performance of our model at an synthon-aggregate level, as we know the pharmacophores and their relative levels of activity.

Lastly, we introduce a new metric to evaluate the quality of our model's predictions by the ability of our model to separately out different classes molecules, which we define as having a better predictor. We first introduce some notation: \textbf{CA-IX} has three distinct groups, $\{g_{para}, g_{meta},g_{other} \}$, in order of protein activity for the para-substituted sulfonamides, meta-substituted sulfonamides, and other molecules respectively. \textbf{HRP} has four distinct groups, $\{g_{e1}, g_{e2}, g_{e3}, g_{other}\}$, in order of protein activity for the three different Michael acceptor electrophiles (described in Figure \ref{fig:synthons}) and other molecules respectively.

From this, we define a multi-class precision-recall area under the curve (PR-AUC) in order to evaluate the ability of our model to differentiate molecule classes. Let $s(g_a | g_b)$ be the computed PR-AUC using $g_a$ as the positive class and $g_b$ as the negative class. Since we know the expected rankings of these molecule classes (ie $g_{para} > g_{meta} > g_{other}$), we can compute the AUC for each pair and then take an unweighted average over all such pairs. Since the data is heavily skewed towards representation of molecules without appreciable activity towards the protein target, we weigh each molecule class equally. These AUC computations are noted exactly in Equations \ref{eq:pr-auc1} and \ref{eq:pr-auc2}:

\begin{equation}
\textrm{PR-AUC}^{\textrm{CA-IX}} = \frac{1}{3} \Big[s(g_{para} | g_{meta}) + s(g_{para} | g_{other}) + s(g_{meta} | g_{other}) \Big],
\label{eq:pr-auc1}
\end{equation}
and 
\begin{equation}
\textrm{PR-AUC}^{\textrm{HRP}} = \frac{1}{6} \Big[s(g_{e1}| g_{e2}) + s(g_{e1} | g_{e3}) + s(g_{e2} | g_{e3}) + s(g_{e1} | g_{other}) + s(g_{e2} | g_{other}) + s(g_{e3} | g_{other})\Big].
\label{eq:pr-auc2}
\end{equation}

\subsection{DEL-Compose captures enrichment of important pharmacophores}

\begin{figure}
    \centering
    \includegraphics[width=0.85\textwidth]{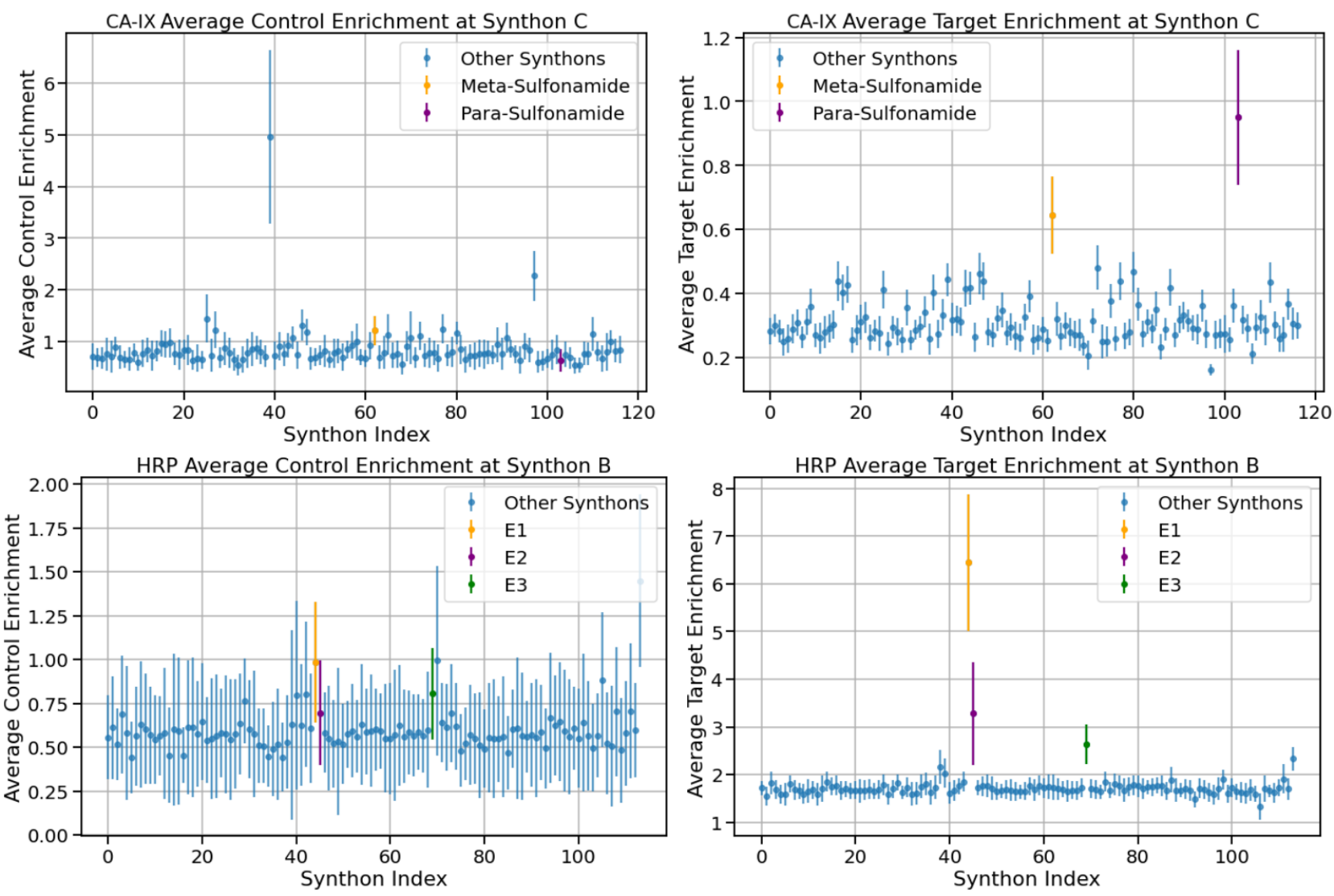}
    \caption{Predicted average marginal enrichment from DEL-Compose of both control and target counts for CA and HRP. Our model can distinguish synthons with high noise in the matrix, but not actual protein binding activity.}
    \label{fig:synthon_marginal}
\end{figure}

Our first result is to show that DEL-Compose can correctly determine the important pharmacophores in the dataset. In Figure \ref{fig:synthon_marginal}, we plot the average marginal enrichment of a synthon as predicted by our model on the test set. For both protein targets, our model correctly enriches the important synthons, which are the benzene-sulfonamides for CA-IX and the Michael-acceptor electrophiles for HRP. Moreover, our model predicts the correct ranking of these different groups. What is particularly interesting to observe is that our model enriches certain non-sulfonamides synthons in the control experiments of CA-IX, but not in the target. This signifies that our model can correctly distinguish the synthons which might have high noise, or off-target matrix binding. 

Next, we compare our deep probabilistic approach to several baselines that compute enrichments based on counts alone. These enrichment baselines try to assess the affinity of a molecule based on some assumed functional form. For instance, the \textbf{Diff Enrichment} score makes the assumption that there is a simple additive effect between the matrix and target counts. Poisson enrichment is taken from~\citet{gerry2019dna}, which computes a maximum likelihood Poisson distribution for the target and matrix counts and then computes a ratio of the target at the lower 95\% confidence interval (CI) and the matrix at the upper 95\% CI. This baseline is perhaps closest to our model, however, our model relates the two distributions directly, whereas this score does not directly correlate the two values. Additionally, we also include the deldenoiser model \cite{komar2020denoising}, which computes a fitness value for each count that correlates to the denoised signal of the counts. We run their model using default parameters, without yield data since there is no yield data for these datasets, and compute enrichment as the average fitness across each target replicate.

\begin{itemize}
\item \textbf{Diff Enrichment}: $\textrm{score} = \frac{1}{n_t}\sum_i c_t^i - \frac{1}{n_m}\sum_j c_m^j $ 

\item \textbf{Ratio Enrichment}: $\textrm{score} = \big[ \big(\frac{1}{n_t}\sum_i c_t^i\big) + 1\big] / \big[\big(\frac{1}{n_m}\sum_j c_m^j $\big)  + 1\big]

\item \textbf{Poisson Enrichment}: $\textrm{score} = \textrm{CI}_{\textrm{lower} 95} [\textrm{Poisson}(\lambda_t)] / \textrm{CI}_{\textrm{upper} 95} [\textrm{Poisson}(\lambda_m)]$
\end{itemize}

Since these baselines are not trained models, but rather explicit functions of the count data, we cannot compare these metrics against our model in terms of predicted likelihood. However, all methods provide a ranking of the test molecules, from which we can compute the aforementioned multi-class PR-AUC. We compare the performance of these models to DEL-Compose on random and synthon splits in Table \ref{tab:abl} and \ref{tabl:abl2} respectively. ``Pre'' and ``Rep'' refer to using the pre-selection and replicate factors in the construction of the output distribution for DEL-Compose. In terms of likelihood, we see that our model that incorporates both pre-selection and replicate factors outperforms all ablations, which validates our intuition that these are important considerations when trying to model DEL data.

Comparing the enrichment baselines to the results of DEL-Compose, we notice that our model outperforms the baselines in terms of multi-class PR-AUC. It is important to note that the baseline metrics do not incorporate the pre-selection data, but even our factorized models without using the pre-selection counts outperform these baselines in most cases. Since the enrichment baselines have oracle access to the actual data, this suggests that DEL-Compose is capturing important aspects of the chemical data. We have not included models learned on top of these computed enrichment scores (which several previous works have proposed), as our model can outperform these oracle metrics already.

\begin{table}
\centerline{
\begin{tabular}{ | c | l | c | c | c | c | }
\hline
 & \textbf{Random Split} & Matrix NLL $\downarrow$ & Target NLL $\downarrow$ & Sum NLL $\downarrow$ & PRC-AUC $\uparrow$\\
 \hline
 \multirow{8}{*}{\rotatebox[origin=c]{90}{\textbf{CA-IX}}} & Diff Enrichment & - & - & - & 0.23 $\pm$ 0.01 \\
 \cline{2-6}
  & Ratio Enrichment & - & - & - & 0.26 $\pm$ 0.02 \\
 \cline{2-6}
 & Poisson Enrichment & - & - & - & 0.25 $\pm$ 0.01 \\
 \cline{2-6}
 & deldenoiser & - & - & - & 0.19 $\pm$ 0.01 \\
 \cline{2-6}
 & DEL-Compose & 3.17 $\pm$ 0.03 & 2.82 $\pm$ 0.04 & 5.99 $\pm$ 0.06 & 0.28 $\pm$ 0.33 \\ 
 \cline{2-6}
 & DEL-Compose (Pre) & 2.97 $\pm$ 0.01 & 2.80 $\pm$ 0.02 & 5.77 $\pm$ 0.02 & \textbf{0.90 $\pm$ 0.03} \\  
 \cline{2-6}
 & DEL-Compose (Rep) & 3.13 $\pm$ 0.03 & 2.65 $\pm$ 0.03  & 5.78 $\pm$ 0.06 & 0.73 $\pm$ 0.34 \\
 \cline{2-6}
 & DEL-Compose (Pre+Rep) & \textbf{2.96 $\pm$ 0.02} & \textbf{2.65 $\pm$ 0.01} & \textbf{5.61 $\pm$ 0.01} & 0.84 $\pm$ 0.07  \\
 \hline
  \multirow{8}{*}{\rotatebox[origin=c]{90}{\textbf{HRP}}} & Diff Enrichment & - & - & - & 0.59 $\pm$ 0.01 \\
 \cline{2-6}
 & Ratio Enrichment & - & - & - & 0.48 $\pm$ 0.01 \\
 \cline{2-6}
 & Poisson Enrichment & - & - & - & 0.57 $\pm$ 0.01 \\
 \cline{2-6}
 & deldenoiser & - & - & - & 0.26 $\pm$ 0.01 \\
 \cline{2-6}
 & DEL-Compose & 6.51 $\pm$ 0.11 & 5.61 $\pm$ 0.09 & 12.12 $\pm$ 0.19 & 0.80 $\pm$ 0.12  \\
 \cline{2-6}
 & DEL-Compose (Pre) & 6.30 $\pm$ 0.03 & 5.35 $\pm$ 0.02 & 11.65 $\pm$ 0.04 & 0.78 $\pm$ 0.03 \\
 \cline{2-6}
 & DEL-Compose (Rep) & 6.39 $\pm$ 0.04 & 5.53 $\pm$ 0.04 & 11.92 $\pm$ 0.08 & \textbf{0.81 $\pm$ 0.04} \\
 \cline{2-6}
 & DEL-Compose (Pre+Rep) & \textbf{6.23 $\pm$ 0.02} & \textbf{5.30 $\pm$ 0.02} & \textbf{11.54 $\pm$ 0.03} & 0.80 $\pm$ 0.04  \\
 \hline
\end{tabular}}
\caption{Metrics for different variants of DEL-Compose compared to baselines on \textbf{random splits}. Metrics are averaged across the test set over 5 different splits.}
\label{tab:abl}
\end{table}

\begin{table}
\centerline{
\begin{tabular}{ | c | l | c | c | c | c | }
\hline
 & \textbf{Synthon Split} & Matrix NLL $\downarrow$ & Target NLL $\downarrow$ & Sum NLL $\downarrow$ & PRC-AUC $\uparrow$ \\
 \hline
 \multirow{8}{*}{\rotatebox[origin=c]{90}{\textbf{CA-IX}}} & Diff Enrichment & - & - & - & 0.22 $\pm$ 0.02 \\
 \cline{2-6}
  & Ratio Enrichment & - & - & - & 0.25 $\pm$ 0.02 \\
 \cline{2-6}
 & Poisson Enrichment & - & - & - & 0.24 $\pm$ 0.02 \\
 \cline{2-6}
  & deldenoiser & - & - & - & 0.19 $\pm$ 0.00 \\
 \cline{2-6}
 & DEL-Compose & 3.58 $\pm$ 0.34 & 2.83 $\pm$ 0.17 & 6.41 $\pm$ 0.46 & 0.13 $\pm$ 0.01 \\ 
 \cline{2-6}
 & DEL-Compose (Pre) & 3.13 $\pm$ 0.13 & 2.81 $\pm$ 0.18 & 5.94 $\pm$ 0.31 & 0.63 $\pm$ 0.42 \\  
 \cline{2-6}
 & DEL-Compose (Rep) & 3.54 $\pm$ 0.29 & 2.64 $\pm$ 0.14 & 6.19 $\pm$ 0.39 & 0.73 $\pm$ 0.33  \\
 \cline{2-6}
 & DEL-Compose (Pre+Rep) & \textbf{3.11 $\pm$ 0.11} & \textbf{2.61 $\pm$ 0.14} & \textbf{5.72 $\pm$ 0.25} & \textbf{0.75 $\pm$ 0.33}  \\
 \hline
  \multirow{8}{*}{\rotatebox[origin=c]{90}{\textbf{HRP}}} & Diff Enrichment & - & - & - & 0.59 $\pm$ 0.01 \\
 \cline{2-6}
 & Ratio Enrichment & - & - & - & 0.48 $\pm$ 0.01 \\
 \cline{2-6}
 & Poisson Enrichment & - & - & - & 0.56 $\pm$ 0.01 \\
 \cline{2-6}
  & deldenoiser & - & - & - & 0.27 $\pm$ 0.00 \\
 \cline{2-6}
 & DEL-Compose & 8.09 $\pm$ 2.04 & 7.36 $\pm$ 1.28 & 15.45 $\pm$ 3.32 & 0.85 $\pm$ 0.16  \\
 \cline{2-6}
 & DEL-Compose (Pre) & 7.44 $\pm$ 2.09 & 6.46 $\pm$ 1.35 & 13.90 $\pm$ 3.44 & 0.85 $\pm$ 0.13  \\
 \cline{2-6}
 & DEL-Compose (Rep) & 8.75 $\pm$ 3.73 & 7.36 $\pm$ 1.69 & 16.11 $\pm$ 5.41 & 0.88 $\pm$ 0.05 \\
 \cline{2-6}
 & DEL-Compose (Pre+Rep) & \textbf{7.29 $\pm$ 2.11} & \textbf{6.26 $\pm$ 1.26} & \textbf{13.55 $\pm$ 3.36} & \textbf{0.90 $\pm$ 0.06}  \\
 \hline
\end{tabular}}
\caption{Metrics for different variants of DEL-Compose compared to baselines on \textbf{synthon splits}. Compared to random splits, the average NLL loss is higher, which confirms our belief that these are more challenging splits of the data for a model to learn over. Our model still outperforms baselines and ablations even on these more challenging splits. }
\label{tabl:abl2}
\end{table}

\subsection{DEL-Compose performs competitively even in low-resource regimes}

\begin{figure}
    \centering
    \includegraphics[width=0.85\textwidth]{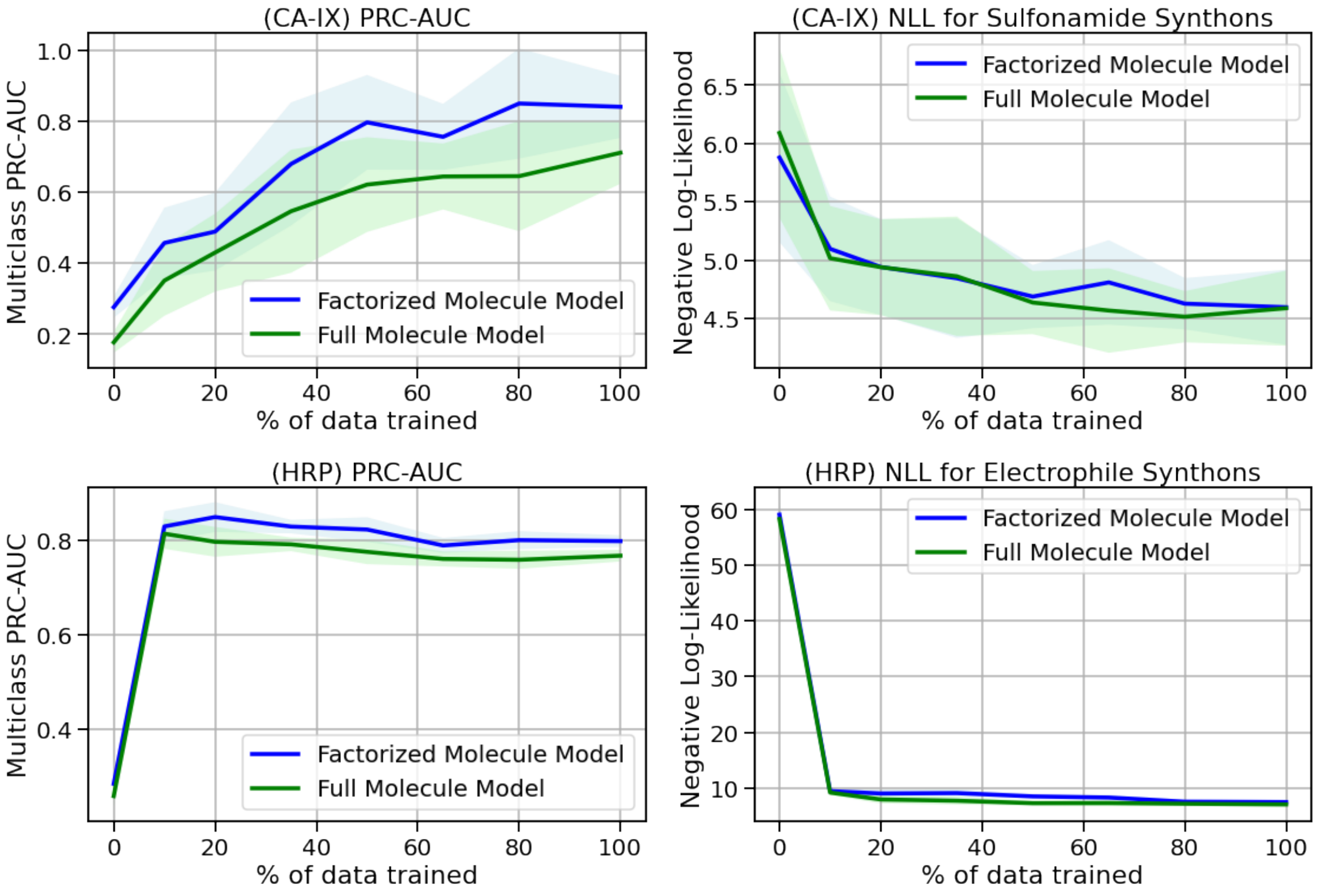}
    \caption{Performance of models using factorized representations vs using full molecule representations. Each model is trained on the same random splits with a different \% of the data heldout. }
    \label{fig:factorized_vs_full}
\end{figure}

One of the main benefits of utilizing a factorized model such as DEL-Compose is that we can avoid building complex enumeration engines for DELs, because we do not require the enumerated full molecule structure. However, while this is beneficial, we want to demonstrate that our factorized models can perform competitively, or even better than models that utilize full molecule representations. To do so, we conduct an in-depth investigation by training both versions of the model under different data-limiting regimes. In Figure \ref{fig:factorized_vs_full}, we compare the performance of both models as a function of amount of data supplied during training. For both CA-IX and HRP, we notice that the multi-class PR-AUC is superior for the factorized model compared to the full model at each point. Meanwhile, the test likelihoods for the two models are very comparable as a function of the amount of data supplied. 

These results support the use of factorized representations as a useful inductive bias to achieve more efficient learning. However, the results are also unsurprising in some sense. The pharmacophores that describe the molecule classes we use are localized within specific synthons, but this property may not hold true for an arbitrary protein. Therefore, we expect that this model might require other improvements or regularizations for more challenging targets and data.

\subsection{{\bf DEL-Compose} facilitates structured interpretation of the data}

\begin{figure}
    \centering
    \includegraphics[width=0.90\textwidth]{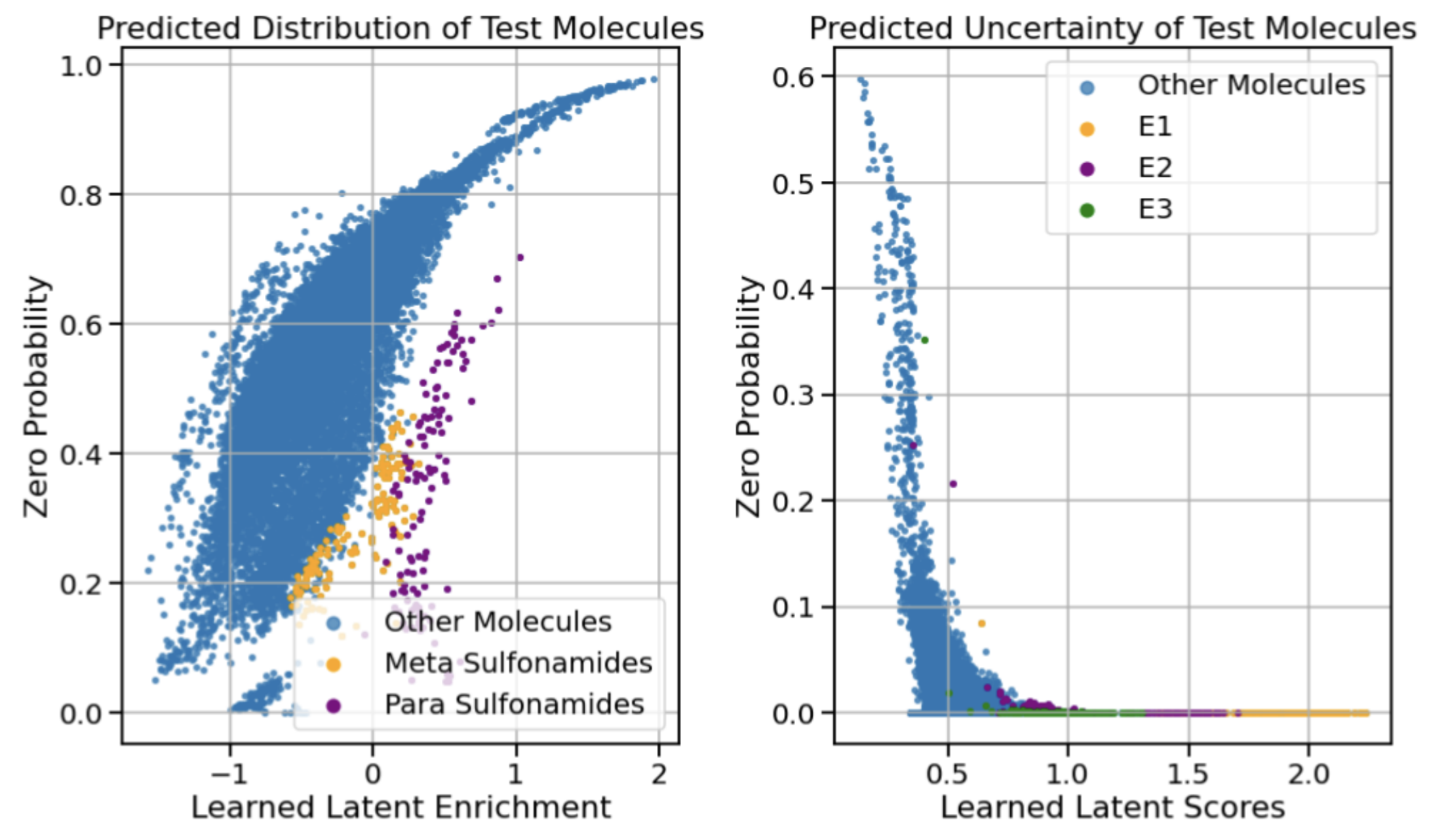}
    \caption{Predicted zero-probability is a good measure of predicted data noise for CA-IX (left) and HRP (right)}
    \label{fig:noise}
\end{figure}

Due to its modeling structure, DEL-Compose provides good interpretability and insights to the model--which is ultimately useful for the chemist using this model. In Figure \ref{fig:noise}, we have plotted the learned latent scores $\lambda$ as a function of the predicted zero probability $p$. For HRP, we see that all the molecules with the known pharmacophores have high predicted scores and low zero-probability--the signal is strong and the noise is low. However, when we turn our attention to the plot for CA-IX, we see that there are a number of molecules with high learned scores, but also high zero-probability--this region of the distribution likely contains more noise. Compared to the HRP data, the predicted scores for the benzene-sulfonamide containing molecules for CA-IX have some uncertainty associated, as implicated by their predicted zero-probabilities. Additionally, our model has nice interpretability with respect to the attention module over the synthons, demonstrating that our model correctly picks out the important synthons.

\begin{figure}
    \centering
    \includegraphics[width=0.92\textwidth]{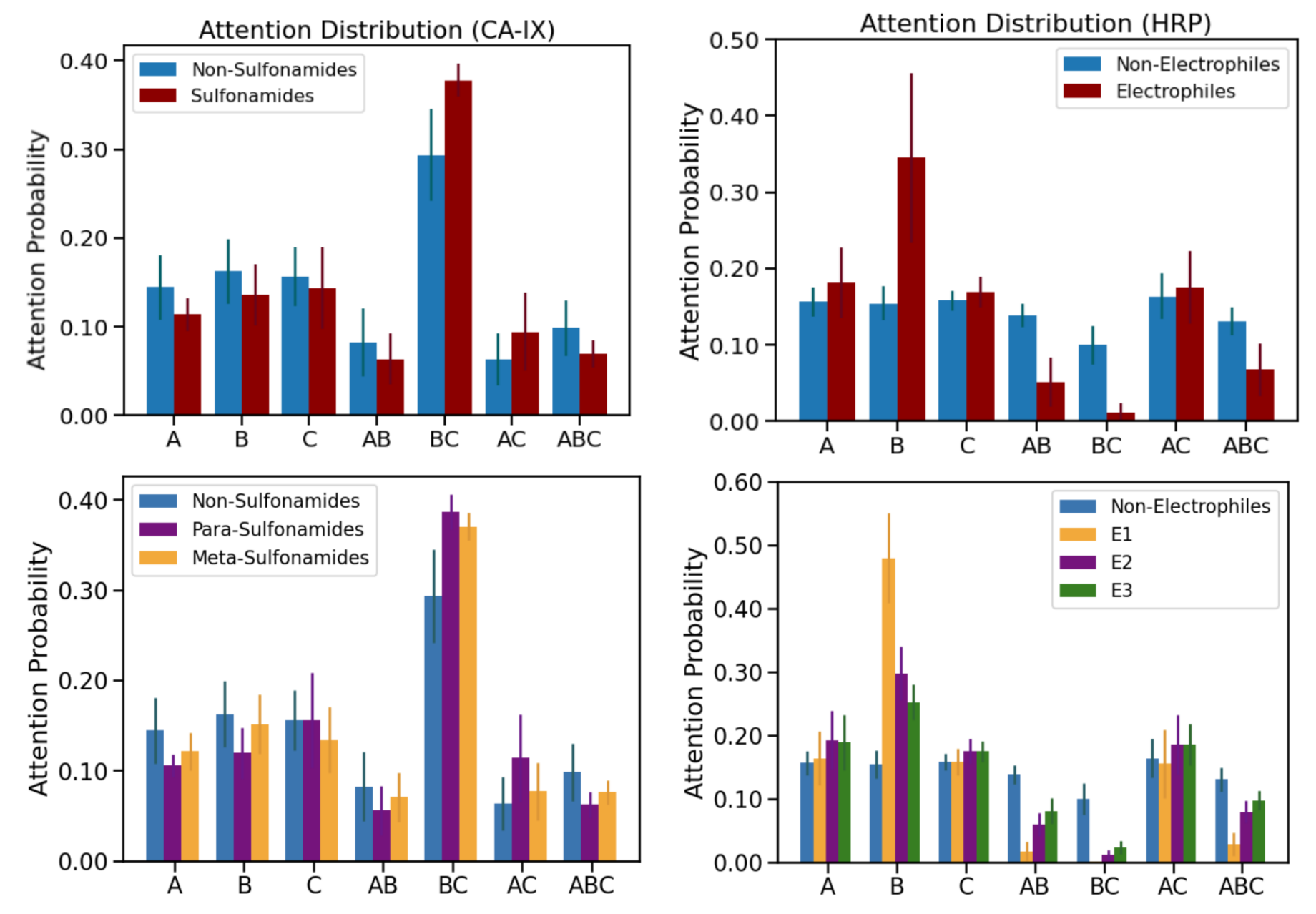}
    \caption{Attention distribution of synthon embeddings. The top set of plots aggregate all the known pharmacophores together, while the bottom set of plots separates out each class of molecules. For CA-IX, the highest attention weights are attributed to the di-synthon $x_{bc}$, while for HRP, the highest attention weights are attributed to the mono-synthon $x_b$.}
    \label{fig:attention}
\end{figure}

In Figure \ref{fig:attention}, we plot the attention distribution over the set of synthon representations of DEL-Compose. The top set of plots aggregate known pharmacophores, while the bottom set of plots delineate individual substructure sets. For CA-IX, we see that the attention probabilities are primarily focused on di-synthon $x_{bc}$, while for HRP, the attention probabilities are primarily distributed on the mono-synthon $x_b$. The sulfonamides are on the C position for CA-IX, while the electrophiles are on the B position for HRP, indicating that both models chose the highest weight to be placed on synthon embeddings incorporating the correct synthon position. Moreover, the weights on the important mono/di-synthons are higher for the molecules with the known pharmacophores versus other molecules in the library. Interestingly, the model for CA-IX chooses the di-synthon $x_{bc}$, instead of just the mono-synthon at the C position. A closer inspection of the enriched di-synthons reveals that the model also highly enriches synthons at the B position which contain sulfonyl chlorides. These synthons form sulfonamides in the tri-synthon structure, albeit internal ones, so the model also predicts that internal sulfonamides have some degree of affinity.

\subsection{Applicability of DEL-Compose}

DEL-Compose offers a structured way to characterize DEL data by decomposing the learned representations from a molecule's synthon composition. Here, we have demonstrated the ability of DEL-Compose to capture important features for in-distribution molecules and to predict enrichments for molecules that align with known active substructures for the proteins in the datasets. Moreover, our model is generally applicable for learning useful representations for many other downstream tasks related to drug discovery. For instance, the representations learned from our model can be used to predict properties of molecules out-of-distribution, or to be used as a guide for generation of new molecules. This contrasts to other works that only try to derive signal from in-library molecules \cite{gerry2019dna, kuai2018randomness, favalli2018dna, komar2020denoising}. Moreover, we present a generative framework that captures the synthesize process of DEL molecules, which does not assume the availability of certain data such as reaction yields. We also demonstrate the viability of training such models without requiring fully enumerated molecule structures.

However, there are also limitations of our experiments here. Due to the nature of the data, our analysis only extends to tasks in which the important chemical substructures are localized within a single synthon, either at the internal B position, or the terminal C position. While we do show that our model attributes meaningful mono- and di-synthons as important, our model can require additional tools to learn powerful representations for more complex targets in order to better capture complex dependency structures. In the proposed DEL-Compose model, we do not consider the explicit chemical structure of di-synthons. However, we can add additional regularizers for the model to predict di-synthon and even tri-synthon structures so that the model can more effectively capture higher order structures. In our work, we also relied on substructure knowledge to split the data into synthon splits. But when such knowledge is unavailable, we can instead either use scaffold splits or cluster the molecules using molecular fingerprints and create splits within each cluster.

\section{Conclusion}

In our work, we proposed a novel method for representing DEL molecules leveraging their combinatorial nature. By incorporating the important experimental factors into our probabilistic model, we demonstrate the ability for our model to pick out the substructures important for particular proteins of interest, CA-IX and HRP. While the model learns useful latent variables that correlate to actual binding properties, we show that our model can also provide interpretable insights for the binding problem. 

\begin{acknowledgement}
We would like to thank Insitro for providing the funding for this project. 

\end{acknowledgement}

\section{Data and Software Availability}

\textbf{Data}: We use publicly available DEL data collected by~\citet{gerry2019dna} on CA-IX and HRP, which is accessible \href{https://pubs.acs.org/doi/pdf/10.1021/jacs.9b01203}{here}. Each protein target has a set of matrix and target experimental replicates. 

\textbf{Software}:  we use \texttt{RDKit} (version \texttt{2020.09.1})~\cite{greg_landrum} to parse our molecules and generate Morgan Fingerprints. The details to replicate our model is in the Methods section, and further training details are provided in the Experiments section. 

\newpage

\begin{suppinfo}

\subsection{CA-IX Enriched Synthons}

From our analysis of our model's attention module, we see that for CA-IX, the model assigns the highest weight to the di-synthon BC, instead of just mono-synthon C, where the terminal benzene-sulfonamides strucutres are located. When we look at which synthons at the B position are enriched, we see that this contains a list of sulfonyl chlorides (see Figure \ref{fig:caix-b}). Full molecules with these synthons have internal sulfonamides, so it is reasonable that the model would upweigh the molecules with these substructures. Interestingly, the internal sulfonamide is not actually apparently when only considering the synthons at the B position alone. 

\begin{figure}
    \centering
    \includegraphics[width=0.95\textwidth]{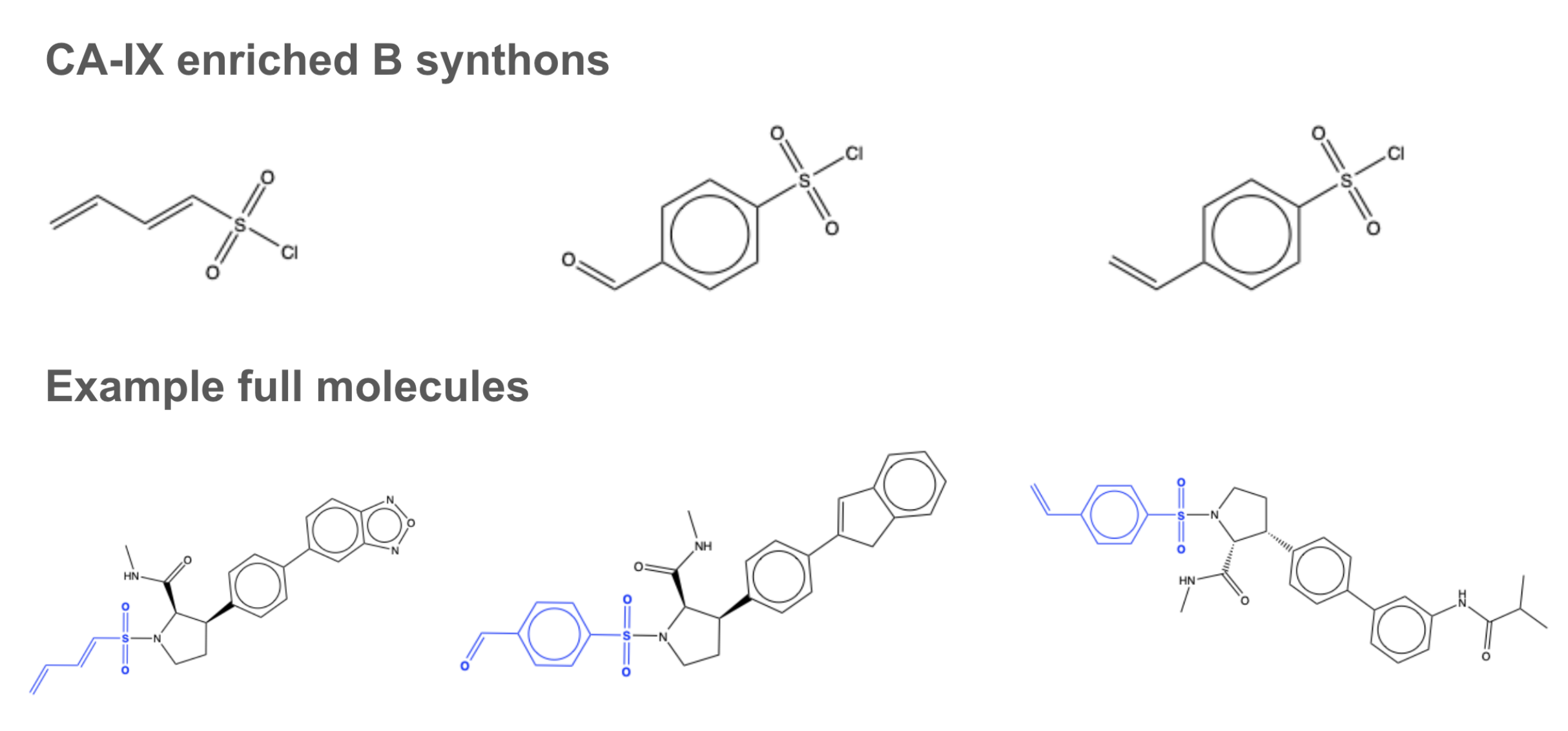}
    \caption{Top enriched synthons at the B position for CA-IX and example full molecules with these synthons (highlighted in blue). }
    \label{fig:caix-b}
\end{figure}

We also see that there are multiple synthons enriched in only the matrix for CA-IX, we have included them in Figure \ref{fig:caix-c} for reference. Our models predict enrichment in the matrix for molecules containing these synthons, but not the enrichment for the protein.

\begin{figure}
    \centering
    \includegraphics[width=0.60\textwidth]{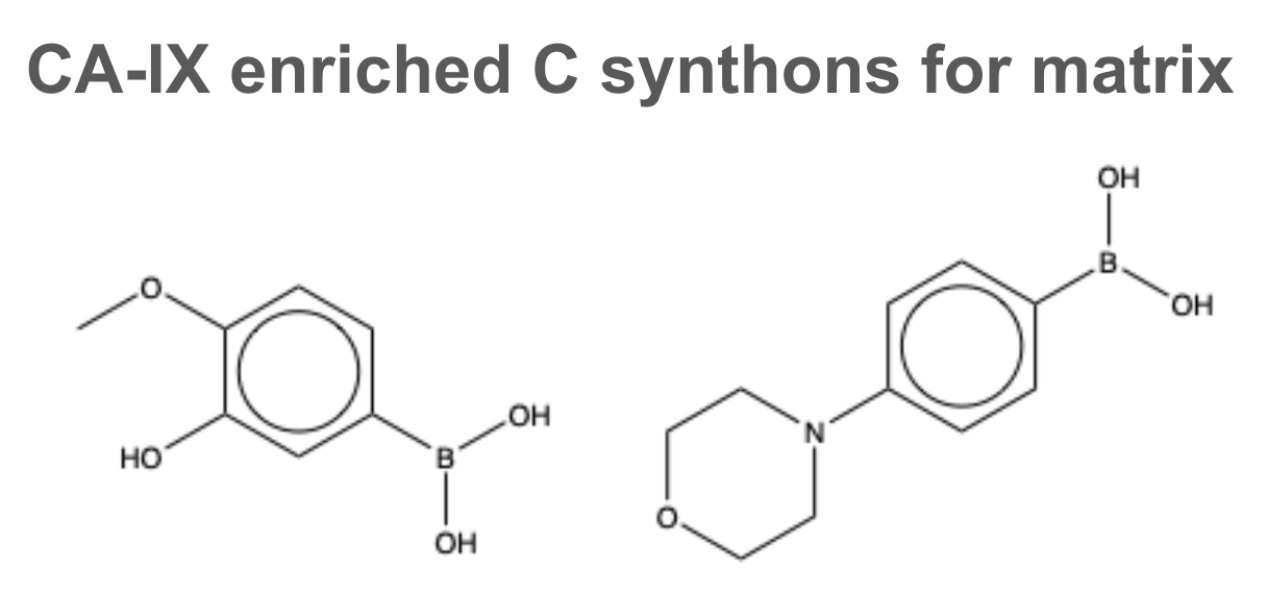}
    \caption{Top enriched synthons at the B position in the matrix data for CA-IX. These molecules were predicted to be enriched in the matrix but not the target by our models. }
    \label{fig:caix-c}
\end{figure}

\end{suppinfo}



\bibliography{references}

\providecommand{\latin}[1]{#1}
\makeatletter
\providecommand{\doi}
  {\begingroup\let\do\@makeother\dospecials
  \catcode`\{=1 \catcode`\}=2 \doi@aux}
\providecommand{\doi@aux}[1]{\endgroup\texttt{#1}}
\makeatother
\providecommand*\mcitethebibliography{\thebibliography}
\csname @ifundefined\endcsname{endmcitethebibliography}  {\let\endmcitethebibliography\endthebibliography}{}
\begin{mcitethebibliography}{29}
\providecommand*\natexlab[1]{#1}
\providecommand*\mciteSetBstSublistMode[1]{}
\providecommand*\mciteSetBstMaxWidthForm[2]{}
\providecommand*\mciteBstWouldAddEndPuncttrue
  {\def\EndOfBibitem{\unskip.}}
\providecommand*\mciteBstWouldAddEndPunctfalse
  {\let\EndOfBibitem\relax}
\providecommand*\mciteSetBstMidEndSepPunct[3]{}
\providecommand*\mciteSetBstSublistLabelBeginEnd[3]{}
\providecommand*\EndOfBibitem{}
\mciteSetBstSublistMode{f}
\mciteSetBstMaxWidthForm{subitem}{(\alph{mcitesubitemcount})}
\mciteSetBstSublistLabelBeginEnd
  {\mcitemaxwidthsubitemform\space}
  {\relax}
  {\relax}

\bibitem[Goodnow~Jr \latin{et~al.}(2017)Goodnow~Jr, Dumelin, and Keefe]{goodnow2017dna}
Goodnow~Jr,~R.~A.; Dumelin,~C.~E.; Keefe,~A.~D. DNA-Encoded Chemistry: Enabling the Deeper Sampling of Chemical Space. \emph{Nat. Rev. Drug Discovery} \textbf{2017}, \emph{16}, 131--147\relax
\mciteBstWouldAddEndPuncttrue
\mciteSetBstMidEndSepPunct{\mcitedefaultmidpunct}
{\mcitedefaultendpunct}{\mcitedefaultseppunct}\relax
\EndOfBibitem
\bibitem[Yuen and Franzini(2017)Yuen, and Franzini]{yuen2017achievements}
Yuen,~L.~H.; Franzini,~R.~M. Achievements, Challenges, and Opportunities in DNA-Encoded Library Research: An Academic Point of View. \emph{ChemBioChem} \textbf{2017}, \emph{18}, 829--836\relax
\mciteBstWouldAddEndPuncttrue
\mciteSetBstMidEndSepPunct{\mcitedefaultmidpunct}
{\mcitedefaultendpunct}{\mcitedefaultseppunct}\relax
\EndOfBibitem
\bibitem[Neri and Lerner(2018)Neri, and Lerner]{neri2018dna}
Neri,~D.; Lerner,~R.~A. DNA-Encoded Chemical Libraries: A Selection System Based on Endowing Organic Compounds With Amplifiable Information. \emph{Annu. Rev. Biochem.} \textbf{2018}, \emph{87}, 479--502\relax
\mciteBstWouldAddEndPuncttrue
\mciteSetBstMidEndSepPunct{\mcitedefaultmidpunct}
{\mcitedefaultendpunct}{\mcitedefaultseppunct}\relax
\EndOfBibitem
\bibitem[Madsen \latin{et~al.}(2020)Madsen, Azevedo, Micco, Petersen, and Hansen]{madsen2020overview}
Madsen,~D.; Azevedo,~C.; Micco,~I.; Petersen,~L.~K.; Hansen,~N. J.~V. An Overview of DNA-Encoded Libraries: A Versatile Tool for Drug Discovery. \emph{Prog. Med. Chem.} \textbf{2020}, \emph{59}, 181--249\relax
\mciteBstWouldAddEndPuncttrue
\mciteSetBstMidEndSepPunct{\mcitedefaultmidpunct}
{\mcitedefaultendpunct}{\mcitedefaultseppunct}\relax
\EndOfBibitem
\bibitem[Satz \latin{et~al.}(2022)Satz, Brunschweiger, Flanagan, Gloger, Hansen, Kuai, Kunig, Lu, Madsen, Marcaurelle, \latin{et~al.} others]{satz2022dna}
Satz,~A.~L.; Brunschweiger,~A.; Flanagan,~M.~E.; Gloger,~A.; Hansen,~N.~J.; Kuai,~L.; Kunig,~V.~B.; Lu,~X.; Madsen,~D.; Marcaurelle,~L.~A.; others DNA-Encoded Chemical Libraries. \emph{Nat. Rev. Methods Primers} \textbf{2022}, \emph{2}, 3\relax
\mciteBstWouldAddEndPuncttrue
\mciteSetBstMidEndSepPunct{\mcitedefaultmidpunct}
{\mcitedefaultendpunct}{\mcitedefaultseppunct}\relax
\EndOfBibitem
\bibitem[Peterson and Liu(2023)Peterson, and Liu]{peterson2023small}
Peterson,~A.~A.; Liu,~D.~R. Small-Molecule Discovery Through DNA-Encoded Libraries. \emph{Nat. Rev. Drug Discovery} \textbf{2023}, 1--24\relax
\mciteBstWouldAddEndPuncttrue
\mciteSetBstMidEndSepPunct{\mcitedefaultmidpunct}
{\mcitedefaultendpunct}{\mcitedefaultseppunct}\relax
\EndOfBibitem
\bibitem[Gironda-Mart{\'\i}nez \latin{et~al.}(2021)Gironda-Mart{\'\i}nez, Donckele, Samain, and Neri]{gironda2021dna}
Gironda-Mart{\'\i}nez,~A.; Donckele,~E.~J.; Samain,~F.; Neri,~D. DNA-Encoded Chemical Libraries: A Comprehensive Review With Succesful Stories and Future Challenges. \emph{ACS Pharmacol. Transl. Sci.} \textbf{2021}, \emph{4}, 1265--1279\relax
\mciteBstWouldAddEndPuncttrue
\mciteSetBstMidEndSepPunct{\mcitedefaultmidpunct}
{\mcitedefaultendpunct}{\mcitedefaultseppunct}\relax
\EndOfBibitem
\bibitem[Reiher \latin{et~al.}(2021)Reiher, Schuman, Simmons, and Wolkenberg]{reiher2021trends}
Reiher,~C.~A.; Schuman,~D.~P.; Simmons,~N.; Wolkenberg,~S.~E. Trends in Hit-To-Lead Optimization Following DNA-Encoded Library Screens. \emph{ACS Med. Chem. Lett.} \textbf{2021}, \emph{12}, 343--350\relax
\mciteBstWouldAddEndPuncttrue
\mciteSetBstMidEndSepPunct{\mcitedefaultmidpunct}
{\mcitedefaultendpunct}{\mcitedefaultseppunct}\relax
\EndOfBibitem
\bibitem[Gerry \latin{et~al.}(2019)Gerry, Wawer, Clemons, and Schreiber]{gerry2019dna}
Gerry,~C.~J.; Wawer,~M.~J.; Clemons,~P.~A.; Schreiber,~S.~L. DNA Barcoding a Complete Matrix of Stereoisomeric Small Molecules. \emph{J. Am. Chem. Soc.} \textbf{2019}, \emph{141}, 10225--10235\relax
\mciteBstWouldAddEndPuncttrue
\mciteSetBstMidEndSepPunct{\mcitedefaultmidpunct}
{\mcitedefaultendpunct}{\mcitedefaultseppunct}\relax
\EndOfBibitem
\bibitem[Kuai \latin{et~al.}(2018)Kuai, O’Keeffe, and Arico-Muendel]{kuai2018randomness}
Kuai,~L.; O’Keeffe,~T.; Arico-Muendel,~C. Randomness in DNA Encoded Library Selection Data Can Be Modeled for More Reliable Enrichment Calculation. \emph{SLAS DISCOVERY} \textbf{2018}, \emph{23}, 405--416\relax
\mciteBstWouldAddEndPuncttrue
\mciteSetBstMidEndSepPunct{\mcitedefaultmidpunct}
{\mcitedefaultendpunct}{\mcitedefaultseppunct}\relax
\EndOfBibitem
\bibitem[Faver \latin{et~al.}(2019)Faver, Riehle, Lancia~Jr, Milbank, Kollmann, Simmons, Yu, and Matzuk]{faver2019quantitative}
Faver,~J.~C.; Riehle,~K.; Lancia~Jr,~D.~R.; Milbank,~J.~B.; Kollmann,~C.~S.; Simmons,~N.; Yu,~Z.; Matzuk,~M.~M. Quantitative Comparison of Enrichment From DNA-Encoded Chemical Library Selections. \emph{ACS Comb. Sci.} \textbf{2019}, \emph{21}, 75--82\relax
\mciteBstWouldAddEndPuncttrue
\mciteSetBstMidEndSepPunct{\mcitedefaultmidpunct}
{\mcitedefaultendpunct}{\mcitedefaultseppunct}\relax
\EndOfBibitem
\bibitem[McCloskey \latin{et~al.}(2020)McCloskey, Sigel, Kearnes, Xue, Tian, Moccia, Gikunju, Bazzaz, Chan, Clark, \latin{et~al.} others]{mccloskey2020machine}
McCloskey,~K.; Sigel,~E.~A.; Kearnes,~S.; Xue,~L.; Tian,~X.; Moccia,~D.; Gikunju,~D.; Bazzaz,~S.; Chan,~B.; Clark,~M.~A.; others Machine Learning on DNA-Encoded Libraries: A New Paradigm for Hit Finding. \emph{J. Med. Chem.} \textbf{2020}, \emph{63}, 8857--8866\relax
\mciteBstWouldAddEndPuncttrue
\mciteSetBstMidEndSepPunct{\mcitedefaultmidpunct}
{\mcitedefaultendpunct}{\mcitedefaultseppunct}\relax
\EndOfBibitem
\bibitem[Zhang \latin{et~al.}(2023)Zhang, Pitman, Dixit, Leelananda, Palacci, Lawler, Belyanskaya, Grady, Franklin, Tilmans, \latin{et~al.} others]{zhang2023building}
Zhang,~C.; Pitman,~M.; Dixit,~A.; Leelananda,~S.; Palacci,~H.; Lawler,~M.; Belyanskaya,~S.; Grady,~L.; Franklin,~J.; Tilmans,~N.; others Building Block-Based Binding Predictions for DNA-Encoded Libraries. \emph{J. Chem. Inf. Model.} \textbf{2023}, \emph{63}, 5120--5132\relax
\mciteBstWouldAddEndPuncttrue
\mciteSetBstMidEndSepPunct{\mcitedefaultmidpunct}
{\mcitedefaultendpunct}{\mcitedefaultseppunct}\relax
\EndOfBibitem
\bibitem[Binder \latin{et~al.}(2022)Binder, Lawler, Grady, Carlson, Leelananda, Belyanskaya, Franklin, Tilmans, and Palacci]{binder2022partial}
Binder,~P.; Lawler,~M.; Grady,~L.; Carlson,~N.; Leelananda,~S.; Belyanskaya,~S.; Franklin,~J.; Tilmans,~N.; Palacci,~H. Partial Product Aware Machine Learning on DNA-Encoded Libraries. \emph{arXiv preprint arXiv:2205.08020} \textbf{2022}, \relax
\mciteBstWouldAddEndPunctfalse
\mciteSetBstMidEndSepPunct{\mcitedefaultmidpunct}
{}{\mcitedefaultseppunct}\relax
\EndOfBibitem
\bibitem[Lim \latin{et~al.}(2022)Lim, Reidenbach, Hua, Mason, Gerry, Clemons, and Coley]{lim2022machine}
Lim,~K.~S.; Reidenbach,~A.~G.; Hua,~B.~K.; Mason,~J.~W.; Gerry,~C.~J.; Clemons,~P.~A.; Coley,~C.~W. Machine Learning on DNA-Encoded Library Count Data Using an Uncertainty-Aware Probabilistic Loss Function. \emph{J. Chem. Inf. Model.} \textbf{2022}, \emph{62}, 2316--2331\relax
\mciteBstWouldAddEndPuncttrue
\mciteSetBstMidEndSepPunct{\mcitedefaultmidpunct}
{\mcitedefaultendpunct}{\mcitedefaultseppunct}\relax
\EndOfBibitem
\bibitem[Ma \latin{et~al.}(2021)Ma, Dreiman, Ruggiu, Riesselman, Liu, James, Sultan, and Koller]{ma2021regression}
Ma,~R.; Dreiman,~G.~H.; Ruggiu,~F.; Riesselman,~A.~J.; Liu,~B.; James,~K.; Sultan,~M.; Koller,~D. Regression Modeling on DNA Encoded Libraries. NeurIPS 2021 AI for Science Workshop. 2021\relax
\mciteBstWouldAddEndPuncttrue
\mciteSetBstMidEndSepPunct{\mcitedefaultmidpunct}
{\mcitedefaultendpunct}{\mcitedefaultseppunct}\relax
\EndOfBibitem
\bibitem[Shmilovich \latin{et~al.}(2023)Shmilovich, Chen, Karaletsos, and Sultan]{shmilovich2023dock}
Shmilovich,~K.; Chen,~B.; Karaletsos,~T.; Sultan,~M.~M. DEL-Dock: Molecular Docking-Enabled Modeling of DNA-Encoded Libraries. \emph{J. Chem. Inf. Model.} \textbf{2023}, \emph{63}, 2719--2727\relax
\mciteBstWouldAddEndPuncttrue
\mciteSetBstMidEndSepPunct{\mcitedefaultmidpunct}
{\mcitedefaultendpunct}{\mcitedefaultseppunct}\relax
\EndOfBibitem
\bibitem[Rogers and Hahn(2010)Rogers, and Hahn]{rogers2010extended}
Rogers,~D.; Hahn,~M. Extended-Connectivity Fingerprints. \emph{J. Chem. Inf. Model.} \textbf{2010}, \emph{50}, 742--754\relax
\mciteBstWouldAddEndPuncttrue
\mciteSetBstMidEndSepPunct{\mcitedefaultmidpunct}
{\mcitedefaultendpunct}{\mcitedefaultseppunct}\relax
\EndOfBibitem
\bibitem[Buller \latin{et~al.}(2011)Buller, Steiner, Frey, Mircsof, Scheuermann, Kalisch, Buhlmann, Supuran, and Neri]{buller2011selection}
Buller,~F.; Steiner,~M.; Frey,~K.; Mircsof,~D.; Scheuermann,~J.; Kalisch,~M.; Buhlmann,~P.; Supuran,~C.~T.; Neri,~D. Selection of Carbonic Anhydrase IX Inhibitors From One Million DNA-Encoded Compounds. \emph{ACS Chem. Biol.} \textbf{2011}, \emph{6}, 336--344\relax
\mciteBstWouldAddEndPuncttrue
\mciteSetBstMidEndSepPunct{\mcitedefaultmidpunct}
{\mcitedefaultendpunct}{\mcitedefaultseppunct}\relax
\EndOfBibitem
\bibitem[Pagnozzi \latin{et~al.}(2022)Pagnozzi, Pala, Biosa, Dallocchio, Dess{\`\i}, Singh, Rogolino, Di~Fiore, De~Simone, Supuran, \latin{et~al.} others]{pagnozzi2022interaction}
Pagnozzi,~D.; Pala,~N.; Biosa,~G.; Dallocchio,~R.; Dess{\`\i},~A.; Singh,~P.~K.; Rogolino,~D.; Di~Fiore,~A.; De~Simone,~G.; Supuran,~C.~T.; others Interaction Studies between Carbonic Anhydrase and a Sulfonamide Inhibitor by Experimental and Theoretical Approaches. \emph{ACS Med. Chem. Lett.} \textbf{2022}, \emph{13}, 271--277\relax
\mciteBstWouldAddEndPuncttrue
\mciteSetBstMidEndSepPunct{\mcitedefaultmidpunct}
{\mcitedefaultendpunct}{\mcitedefaultseppunct}\relax
\EndOfBibitem
\bibitem[Gan \latin{et~al.}(2013)Gan, Kaminska, Yang, Liew, Leow, So, Tu, Roy, Yap, Kang, \latin{et~al.} others]{gan2013identification}
Gan,~F.-F.; Kaminska,~K.~K.; Yang,~H.; Liew,~C.-Y.; Leow,~P.-C.; So,~C.-L.; Tu,~L.~N.; Roy,~A.; Yap,~C.-W.; Kang,~T.-S.; others Identification of Michael Acceptor-Centric Pharmacophores With Substituents That Yield Strong Thioredoxin Reductase Inhibitory Character Correlated to Antiproliferative Activity. \emph{Antioxid. Redox Signaling} \textbf{2013}, \emph{19}, 1149--1165\relax
\mciteBstWouldAddEndPuncttrue
\mciteSetBstMidEndSepPunct{\mcitedefaultmidpunct}
{\mcitedefaultendpunct}{\mcitedefaultseppunct}\relax
\EndOfBibitem
\bibitem[Kingma and Ba(2014)Kingma, and Ba]{kingma2014adam}
Kingma,~D.~P.; Ba,~J. Adam: A Method for Stochastic Optimization. \emph{arXiv preprint arXiv:1412.6980} \textbf{2014}, \relax
\mciteBstWouldAddEndPunctfalse
\mciteSetBstMidEndSepPunct{\mcitedefaultmidpunct}
{}{\mcitedefaultseppunct}\relax
\EndOfBibitem
\bibitem[Gawlikowski \latin{et~al.}(2023)Gawlikowski, Tassi, Ali, Lee, Humt, Feng, Kruspe, Triebel, Jung, Roscher, \latin{et~al.} others]{gawlikowski2023survey}
Gawlikowski,~J.; Tassi,~C. R.~N.; Ali,~M.; Lee,~J.; Humt,~M.; Feng,~J.; Kruspe,~A.; Triebel,~R.; Jung,~P.; Roscher,~R.; others A Survey of Uncertainty in Deep Neural Networks. \emph{Artificial Intelligence Review} \textbf{2023}, 1--77\relax
\mciteBstWouldAddEndPuncttrue
\mciteSetBstMidEndSepPunct{\mcitedefaultmidpunct}
{\mcitedefaultendpunct}{\mcitedefaultseppunct}\relax
\EndOfBibitem
\bibitem[Heid \latin{et~al.}(2023)Heid, McGill, Vermeire, and Green]{heid2023characterizing}
Heid,~E.; McGill,~C.~J.; Vermeire,~F.~H.; Green,~W.~H. Characterizing Uncertainty in Machine Learning for Chemistry. \emph{J. Chem. Inf. Model.} \textbf{2023}, \relax
\mciteBstWouldAddEndPunctfalse
\mciteSetBstMidEndSepPunct{\mcitedefaultmidpunct}
{}{\mcitedefaultseppunct}\relax
\EndOfBibitem
\bibitem[Wu \latin{et~al.}(2018)Wu, Ramsundar, Feinberg, Gomes, Geniesse, Pappu, Leswing, and Pande]{wu2018moleculenet}
Wu,~Z.; Ramsundar,~B.; Feinberg,~E.~N.; Gomes,~J.; Geniesse,~C.; Pappu,~A.~S.; Leswing,~K.; Pande,~V. MoleculeNet: A Benchmark for Molecular Machine Learning. \emph{Chem. Sci.} \textbf{2018}, \emph{9}, 513--530\relax
\mciteBstWouldAddEndPuncttrue
\mciteSetBstMidEndSepPunct{\mcitedefaultmidpunct}
{\mcitedefaultendpunct}{\mcitedefaultseppunct}\relax
\EndOfBibitem
\bibitem[K{\'o}m{\'a}r and Kalinic(2020)K{\'o}m{\'a}r, and Kalinic]{komar2020denoising}
K{\'o}m{\'a}r,~P.; Kalinic,~M. Denoising DNA Encoded Library Screens With Sparse Learning. \emph{ACS Combinatorial Science} \textbf{2020}, \emph{22}, 410--421\relax
\mciteBstWouldAddEndPuncttrue
\mciteSetBstMidEndSepPunct{\mcitedefaultmidpunct}
{\mcitedefaultendpunct}{\mcitedefaultseppunct}\relax
\EndOfBibitem
\bibitem[Favalli \latin{et~al.}(2018)Favalli, Bassi, Scheuermann, and Neri]{favalli2018dna}
Favalli,~N.; Bassi,~G.; Scheuermann,~J.; Neri,~D. DNA-Encoded Chemical Libraries–Achievements and Remaining Challenges. \emph{FEBS Lett.} \textbf{2018}, \emph{592}, 2168--2180\relax
\mciteBstWouldAddEndPuncttrue
\mciteSetBstMidEndSepPunct{\mcitedefaultmidpunct}
{\mcitedefaultendpunct}{\mcitedefaultseppunct}\relax
\EndOfBibitem
\bibitem[Landrum \latin{et~al.}(2021)Landrum, Tosco, Kelley, sriniker, gedeck, Ric, Vianello, Schneider, Dalke, and N]{greg_landrum}
Landrum,~G.; Tosco,~P.; Kelley,~B.; sriniker; gedeck; Ric; Vianello,~R.; Schneider,~N.; Dalke,~A.; N,~D. rdkit/rdkit: 2021\_09\_4 (Q3 2021) Release. 2021; \url{https://doi.org/10.5281/zenodo.5835217}, Accessed 2023-12-04\relax
\mciteBstWouldAddEndPuncttrue
\mciteSetBstMidEndSepPunct{\mcitedefaultmidpunct}
{\mcitedefaultendpunct}{\mcitedefaultseppunct}\relax
\EndOfBibitem
\end{mcitethebibliography}





\end{document}